\documentstyle[12pt]{article}

\batchmode
  \newfont{\footscrfont}{rsfs10}
  \newfont{\footbbbfont}{msbm10}
\errorstopmode

\newif\ifscrf\scrftrue
\ifx\footscrfont\nullfont
  \scrffalse
\fi

\newif\ifamsf\amsftrue
\ifx\footbbbfont\nullfont
  \amsffalse
\fi

\def\ppnumber{\vbox{\baselineskip16pt\hbox{DUK-TH-94-70}
  \hbox{IASSNS-HEP-94/37}}}
\def\ppdate{May, 1994}
\def\pplogo{\vbox{\kern-\headheight\kern -17pt
\halign{##&##\hfil\cr&{
\ppnumber}\cr\rule{0pt}{2.5ex}&\ppdate\cr}
}}

\makeatletter
\date{}
\def\dedicatory#1{\def\@date{\normalsize\it#1}}
\def\subjclass#1{\def\@thefnmark{}\@footnotetext{1991
    {\it Mathematics Subject Classification.} #1}}
\def\keywords#1{\def\@thefnmark{}\@footnotetext{
    {\it Key words and phrases.} #1}}

\def\ps@firstpage{\ps@empty \def\@oddhead{\hss\pplogo}%
  \let\@evenhead\@oddhead 
}
\def\maketitle{\par
 \begingroup
 \def\thefootnote{\fnsymbol{footnote}}
 \def\@makefnmark{\hbox
 to 0pt{$^{\@thefnmark}$\hss}}
 \if@twocolumn
 \twocolumn[\@maketitle]
 \else \newpage
 \global\@topnum\z@ \@maketitle \fi\thispagestyle{firstpage}\@thanks
 \endgroup
 \setcounter{footnote}{0}
 \let\maketitle\relax
 \let\@maketitle\relax
 \gdef\@thanks{}\gdef\@author{}\gdef\@title{}\let\thanks\relax}

\def\abstract{\if@twocolumn
\section*{Abstract}
\else \small
\begin{center}
{\bf ABSTRACT}
\end{center}
\quotation
\fi}

\def\thebibliography#1{\section*{References\@mkboth
 {REFERENCES}{REFERENCES}}\small\list
 {[\arabic{enumi}]}{\settowidth\labelwidth{[#1]}\leftmargin\labelwidth
 \advance\leftmargin\labelsep
 \usecounter{enumi}}
 \def\newblock{\hskip .11em plus .33em minus .07em}
 \sloppy\clubpenalty4000\widowpenalty4000
 \sfcode`\.=1000\relax}

\newif\iffn\fnfalse

\@ifundefined{reset@font}{\let\reset@font\empty}{} 
\long\def\@footnotetext#1{\insert\footins{\reset@font\footnotesize
    \interlinepenalty\interfootnotelinepenalty
    \splittopskip\footnotesep
    \splitmaxdepth \dp\strutbox \floatingpenalty \@MM
    \hsize\columnwidth \@parboxrestore
   \edef\@currentlabel{\csname p@footnote\endcsname\@thefnmark}\@makefntext
    {\rule{\z@}{\footnotesep}\ignorespaces
      \fntrue#1\fnfalse\strut}}}
\makeatother

\ifamsf
  \newfont{\bigbbbfont}{msbm10 scaled\magstep2}
  \newfont{\bbbfont}{msbm10 scaled\magstep1}  
  \newfont{\smallbbbfont}{msbm8}
  \newfont{\tinybbbfont}{msbm6}
  \newfont{\smallfootbbbfont}{msbm7}
  \newfont{\tinyfootbbbfont}{msbm5}
\fi

\ifscrf
  \newfont{\scrfont}{rsfs10 scaled\magstep1}  
  \newfont{\smallscrfont}{rsfs7}
  \newfont{\tinyscrfont}{rsfs7}
  \newfont{\smallfootscrfont}{rsfs7}
  \newfont{\tinyfootscrfont}{rsfs7}
\fi

\ifamsf
  \newcommand{\Bbb}[1]{\iffn
      \mathchoice{\mbox{\footbbbfont #1}}{\mbox{\footbbbfont #1}}
      {\mbox{\smallfootbbbfont #1}}{\mbox{\tinyfootbbbfont #1}}\else
      \mathchoice{\mbox{\bbbfont #1}}{\mbox{\bbbfont #1}}
      {\mbox{\smallbbbfont #1}}{\mbox{\tinybbbfont #1}}\fi}
\else
  \def\bigbbbfont{\bf}
  \def\Bbb{\bf}
\fi

\ifscrf
  \newcommand{\Scr}[1]{\iffn
    \mathchoice{\mbox{\footscrfont #1}}{\mbox{\footscrfont #1}}
    {\mbox{\smallfootscrfont #1}}{\mbox{\tinyfootscrfont #1}}\else
    \mathchoice{\mbox{\scrfont #1}}{\mbox{\scrfont #1}}
    {\mbox{\smallscrfont #1}}{\mbox{\tinyscrfont #1}}\fi}
\else
  \def\Scr{\cal}
\fi

\def\operatorname#1{\mathop{\rm #1}\nolimits}
\def\C{{\Bbb C}}

\def\P{{\Bbb P}}

\def\Z{{\Bbb Z}}

\def\Hom{\operatorname{Hom}}

\def\opeq#1{\advance\lineskip#1 \advance\baselineskip#1
	\advance\lineskiplimit#1}
\def\eqalign#1{\null\,\vcenter{\opeq{2.5\jot}\mathsurround=0pt
	\everycr={}\tabskip=0pt
	\halign{\strut\hfil$\displaystyle{##}$&$\displaystyle{{}##}$\hfil
	\crcr#1\crcr}}\,\null}

\def\sm{$\sigma$-model}
\def\nlsm{non-linear \sm}

\def\CY{Calabi-Yau}

\def\cM{{\Scr M}}

\def\cMc{{\hfuzz=100cm\hbox to 0pt{$\;\overline{\phantom{X}}$}\cM}}

\def\ff#1#2{{\textstyle\frac{#1}{#2}}}

\newcommand{\text}[1]{\mathchoice{\mbox{\rm #1}}{\mbox{\rm #1}}
    {\mbox{\scriptsize\rm #1}}{\mbox{\tiny\rm #1}}}

\ifscrf
  
\else
  
\fi

\ifamsf
  \def\ltimes{\mathbin{\mbox{\bbbfont\char"6E}}}
  \def\nmid{\mathbin{\mbox{\bbbfont\char"2D}}}
\else
  \def\ltimes{.}
  \def\nmid{\hbox{does not divide}}
\fi

\makeatletter
\relax
\@writefile{toc}{\string\contentsline\space {section}{\string\numberline\space
{1}Introduction}{1}}
\@writefile{toc}{\string\contentsline\space {section}{\string\numberline\space
{2}A-model Analysis}{2}}
\@writefile{toc}{\string\contentsline\space {section}{\string\numberline\space
{3}The B-model of the Mirror}{4}}
\@writefile{toc}{\string\contentsline\space {section}{\string\numberline\space
{4}Some curve counting}{6}}
\@writefile{toc}{\string\contentsline\space {section}{\string\numberline\space
{5}Conclusions}{9}}
\makeatother

\begin{document}
\setcounter{page}0
\title{\LARGE Chiral Rings Do Not Suffice:\\
$N$=(2,2) Theories with Nonzero\\
Fundamental Group\\[10mm]
}
\author{
Paul S. Aspinwall,\\[0.7cm]
\normalsize School of Natural Sciences,\\
\normalsize Institute for Advanced Study,\\
\normalsize Princeton, NJ  08540\\[10mm]
David R. Morrison \\[0.7cm]
\normalsize  Department of Mathematics, \\
\normalsize  Box 90320, \\
\normalsize  Duke University, \\
\normalsize  Durham, NC 27708-0320}

{\hfuzz=10cm\maketitle}

\def\Large{\large}
\def\LARGE{\large\bf}


\begin{abstract}

The K\"ahler moduli space of a particular
non-simply-connected Calabi-Yau
manifold is mapped out using mirror symmetry. It is found that, for
the model considered, the chiral ring may be identical for different
associated conformal field theories. This ambiguity is explained in
terms of both A-model and B-model language.
It also provides an apparent counterexample to the global Torelli problem
for Calabi-Yau threefolds.

\end{abstract}

\vfil\break

\section{Introduction}		\label{s:intro}

One of the very first \CY\ spaces to be considered for a superstring
compactification was defined as follows \cite{CHSW:}. Take the complex
projective space $\P^4$ with homogeneous coordinates
$[z_0,z_1,\ldots,z_4]$ and construct a quintic hypersurface within it
{}from the condition
\begin{equation}
  p=z_0^5+z_1^5+z_2^5+z_3^5+z_4^5=0.	\label{eq:Fermat}
\end{equation}
Let us denote this space by $Q$. Now define a \CY\ space as
 $X=Q/G$ where $G$ is the freely acting group of identifications
isomorphic to $\Z_5\times\Z_5$ generated by
\begin{equation}
  \eqalign{g_1:[z_0,z_1,z_2,z_3,z_4]&\mapsto
	[z_0,\zeta z_1,\zeta^2z_2,\zeta^3z_3,\zeta^4z_4],
	\qquad \zeta=e^{2\pi i/5},\cr
	g_2:[z_0,z_1,z_2,z_3,z_4]&\mapsto
	[z_1,z_2,z_3,z_4,z_0].}	\label{eq:Ggen}
\end{equation}
The smooth \CY\ manifold $X$ has Euler characteristic $-8$ and can
thus be used to build a four generation model. Although these days,
four generations is not considered an attractive feature of a model, it
will still prove interesting to study the space $X$ and, in
particular, the moduli space of $N$=(2,2) superconformal field theories
containing a point corresponding to this manifold.

$N$=(2,2) superconformal field theories are used to represent a \CY\
compactification with the ``spin connection embedded in the gauge
group'' \cite{CHSW:}. The moduli space of such conformal field
theories, i.e., the moduli space of string vacua of this type is
actually quite simple to analyze in many cases. One of the reasons for
this is the existence of the ``chiral ring'' (see, for
example, \cite{LVW:}). To analyze a generic conformal field theory it
is usual to concentrate on the set of primary fields, which unfortunately are
infinite in number.\footnote{If the conformal
field theory is rational, the number of primary fields
will be finite and the analysis
may not be so difficult. However, it is believed that such rational conformal
field theories do not deform and thus can, at best, form only a
countable dense subset of the space of conformal field theories
(like the set of rational numbers in the space of
reals).} However, in the case of $N$=(2,2)
superconformal field theories we may look at two distinguished
 subsets of the primary
fields, the {\em chiral\/} primary fields and the {\em antichiral\/}
primary fields.  (The difference between the definitions
is simply a change in the sign of the $U(1)$ charge). The advantage of this is
that the number of (anti)chiral primary fields is finite, and remains constant
over the moduli space. This makes analysis of the set of such fields much
easier.  Furthermore, the ``na\"{\i}ve product''
\begin{equation}
(\phi\chi)(z)=\lim_{z'\to z}\phi(z')\chi(z)
\end{equation}
of two chiral
primary fields is again a chiral primary field, giving the set of such
fields the structure of a ring.  The structure constants of this ring
can be found by calculating
 two point and  three point functions.

An $N$=(2,2) superconformal field theory may be twisted to form a
topological field theory. This may be done in two inequivalent ways
\cite{W:AB} yielding the ``A-model'' or ``B-model''. The observables
in each of these models correspond to the (anti)chiral primary fields of the
original $N$=(2,2) theory, the difference amounting to a choice of
sign of one of the $U(1)$ charges in the $N$=(2,2) superconformal
algebra. Deformations of the $N$=(2,2) superconformal field theory can
generated by (anti)chiral primary fields and thus map into
deformations of the A-model
and B-model, where they are detected by the ring structures.
It is tempting to speculate that analyzing the
two point and three
point functions of the observables in the two topological field
theories gives sufficient information to classify completely the
$N$=(2,2) theory. That is, the two chiral rings of the conformal
field theory might contain sufficient information to obtain all other
correlation functions from them.

A simple counterexample to this proposition is the
complex 3-torus \cite{Vafa:vacua}. In
this case the A-model (having no instantons at tree-level) is too
trivial and contains no local
information. One possible explanation for this is that
in the case of the torus,
 the local supersymmetry is actually $N$=(4,4). Whenever this
happens the moduli space takes on a significantly different form which
no longer splits naturally into A-model and B-model part (see, for
example, \cite{N:torus,AM:K3p}). One might thus modify the proposal to
apply to theories with only $N$=(2,2) symmetry. We shall see from our
example that this too fails---for global rather than local reasons.

In section \ref{s:A} we will discuss the A-model on $X$. In section
\ref{s:B} the moduli space will be formulated by considering the
B-model on the mirror of $X$. To help understand the form of the
moduli space we will consider rational and elliptic curves on $X$ in
section \ref{s:count} and finally in section \ref{s:conc} we will
present concluding remarks.


\section{A-model Analysis}   	\label{s:A}

In terms of the geometric data of the target space, the B-model
captures the information concerned with the complex structure. The
A-model moduli however are concerned with variation of the
complexified K\"ahler form. By expressing the constraint in the form
(\ref{eq:Fermat}) we have effectively fixed the complex structure and
thus frozen the B-model data. Thus we will concentrate only on the
A-model.

In the conventional approach to the A-model one describes a \nlsm\
whose target space is equipped with a K\"ahler metric
given by a K\"ahler form $J$ and
a real 2-form, $B$. These may be combined to form a complex 2-form
$B+iJ$ upon whose cohomology class the A-model depends. This
description does not capture all the possibilities however.

For a map $\phi:\Sigma\to X$ from the world-sheet into the target
space, the A-model correlation functions vary as a function of
\begin{equation}
  \xi = \exp(2\pi i\int_\Sigma \phi^*(B+iJ)),
\end{equation}
where $\phi^*$ is the pull-back. Thus the degree of freedom
represented by $B+iJ$ can be thought of as an element of
$\Hom(H_2(X),\C^*)$. That is, it associates some non-zero complex
number, $\xi$, to each homology class of the image of $\Sigma$ in
$X$. The group $\Hom(H_2(X),\C^*)$ may contain more freedom than that
described by the complex 2-form
$B+iJ$. It is possible that the singular homology group
$H_2(X)$ contains {\em torsion}, i.e., there is an element $\tau\in
H_2(X)$ such that $N\tau\sim0$ for some integer $N$. Using only de
Rham cohomology in the form of $B+iJ$ will then miss the corresponding
torsion elements of $\Hom(H_2(X),\C^*)$. Since it would seem natural
to allow for torsion elements of $H_2(X)$ to be associated with a
non-trivial $\xi$ it would appear that the A-model moduli space is
better described as $\Hom(H_2(X),\C^*)$ rather than the potentially
smaller space of $B+iJ$'s.
This is closely related to issues studied in \cite{DW:gp}.
This form of the moduli space can also be justified by looking at
fundamental properties of maps of the world-sheet into the target space
\cite{DW:gp,SW:spin,Vafa:tor}.

It is important to note that the A-model picture presented here
depends on which ``phase'' of moduli space we are in in the sense of
\cite{W:phase,AGM:II}. We will assume that we are in some
neighbourhood of the large radius limit of $X$ and thus in the \CY\
phase. When one leaves this phase, the moduli space will no longer
appear to be in the form $\Hom(H_2(X),\C^*)$.

For the quintic threefold, $Q$, $\dim H^2_{\rm DR}(Q)=1$ and there is
no torsion in $H_2(Q)$. Thus, in the neighbourhood of the large radius
limit (where the A-model is well-defined) the moduli space is locally
isomorphic to $\Hom(H_2(Q),\C^*)\cong \C^*$. Since $Q$ is a simply-connected
non-ramified cover of $X$, we have that
$\pi_2(X)\cong\pi_2(Q)\cong H_2(Q)\cong\Z$ \cite{BT:}.
Since $X=Q/G$ we also have that $\pi_1(X) \cong G\cong \Z_5\times\Z_5$.
Given $\pi_1(X)$ and $\pi_2(X)$ we may calculate $H_2(X)$ and consequently
$\Hom(H_2(X),\C^*)\cong H^2(X,\C^*)$ by the method of Eilenberg and MacLane
\cite{EM:hom}.
According to \cite{EM:hom} there is an exact sequence
\begin{equation}
  0\to \pi_2(X) \to H_2(X) \to H_2(G) \to 0
\label{eq:ES2}
\end{equation}
which relates the homotopy and homology groups of $X$ to the group
homology $H_2(G)$.  In the present case, the group homology can
be calculated as
$H_2(\Z_5\times\Z_5)\cong\Z_5$, and the exact sequence (\ref{eq:ES2})
becomes
\begin{equation}
  0\to\Z\to H_2(X)\to\Z_5\to 0.	\label{eq:ES3}
\end{equation}
There are two possibilities for $H_2(X)$ compatible with (\ref{eq:ES3}),
depending on whether the exact sequence splits:
either $H_2(X)\cong\Z\times\Z_5$, or $H_2(X)\cong\Z$.  In either case
there will be a homology class $e$ which is not represented by
a sphere; in the former case, $e$ may be chosen so that $5e\sim0$
and in the latter so that $5e$ generates $\pi_2(X)\cong H_2(Q)$.
(These two possibilities are mutually exclusive.)

In fact, it is the second possibility $H_2(X)\cong\Z$ which occurs for
our example.  We will show this in section 4 by exhibiting an elliptic
curve $E$ on $X$ whose inverse image $\pi^{-1}(E)$ on $Q$ is an irreducible
elliptic curve of degree $5$.  The homology class $e$ of $E$ cannot
lie in $\pi_2(X)$, since for every rational curve $\Gamma$ on $X$,
the inverse image $\pi^{-1}(\Gamma)$ consists of 25 disjoint rational
curves, all of the same degree $d$, so that the degree of $\pi^{-1}(\Gamma)$
is a multiple of 25.

The group $H^2(X,\C^*)\cong\Hom(H_2(X),\C^*)$
which describes the degree of freedom
represented by $B+iJ$ fits in an exact sequence of its own
\begin{equation}
  0\to\Z_5\to H^2(X,\C^*)\mathrel{\mathop\to^\alpha}\C^*\to 0,	\label{eq:ES1}
\end{equation}
whose structure is easily deduced from that of (\ref{eq:ES3}).
In particular, $H^2(X,\C^*)\cong\C^*$ and
the map $\alpha$ is a five-fold cover.

This
occurrence of $H^2(G,\C^*)$ when modding out a space by a group action
$G$ was first discussed in the context of string theory in
\cite{Vafa:tor}\footnote{In that paper only the
$B$-field is discussed leading to the equivalent group $H^2(G,U(1))$.}
where it was given the name ``discrete torsion''. Note that in our
example this name is somewhat misleading since $H^2(X,\C^*)$ is torsion
free.


\section{The B-model of the Mirror}	\label{s:B}

Given $X$ we now hope to find another space $Y$, which is the ``mirror'' of
$X$, such that the A-model on $X$ is equivalent to the B-model on $Y$.
To be more precise, $Y$ will actually be in the Landau-Ginzburg phase
rather than \CY\ phase but the phase picture is not important for the
B-model and we may imagine $Y$ to be a \CY\ manifold for all practical
purposes.
In the case of the quintic and many of its quotients the mirror is
given by the method of \cite{GP:orb}. $X$ does not quite
fall into this class but we may use an extension of this
method \cite{ALR:coup}
to find the mirror. Define  $G_1$, isomorphic to
$\Z_5$, to be the group generated by the element $g_1$  defined in
eq.~(\ref{eq:Ggen}). By the
method of \cite{GP:orb},
the manifold $Q/G_1$ is
known to be mirror to $Q/\widetilde{G_1}$, where
$\widetilde{G_1}\cong\Z_5\times\Z_5$ is
generated by $g_1$ and another element $g_3$ defined by
\begin{equation}
  g_3:[z_0,z_1,z_2,z_3,z_4]\mapsto
	[z_0,\zeta z_1,\zeta^3z_2,\zeta z_3,z_4],
	\qquad \zeta=e^{2\pi i/5}.
\end{equation}
$g_2$ now acts on the mirror pair of theories corresponding to the
spaces $Q/G_1$ and
$Q/\widetilde{G_1}$ in precisely the same manner
 (once the mirror map is taken into
account). Thus we may divide both spaces (i.e., orbifold both
conformal field theories) by the group generated by this action to
yield another mirror pair. This pair consists of $X$ and $Y\cong
Q/\widetilde{G}$ where $\widetilde{G}\cong\Z_5\ltimes(\Z_5)^2$ and is
generated by $g_1$, $g_2$ and $g_3$.

The moduli space of the B-model consists of varying the complex
structure of $Y$ which may be done by varying the defining equation
(\ref{eq:Fermat}) of $Q$.
There is only one deformation compatible with
the group $\widetilde{G}$ and we follow
\cite{CDGP:} by using the following parametrization
\begin{equation}
  p_\psi=z_0^5+z_1^5+z_2^5+z_3^5+z_4^5-5\psi z_0z_1z_2z_3z_4=0.
		\label{eq:Bmod}
\end{equation}

Further analysis of the moduli space and the mirror map between this
B-model and the A-model of section \ref{s:A} is now very close to the
analysis of \cite{CDGP:} where the A-model considered was that
associated to $Q$. The only difference between the B-model considered
here and that of \cite{CDGP:} is that a different 125 element group of
symmetries (call it $\widehat G$)
is used to divide $Q$, although both lead to the same
general form of defining equation (\ref{eq:Bmod}). Little of the
analysis of \cite{CDGP:} depends on the exact form of the group
dividing $Q$ and so can be copied over to the case considered here.

In \cite{CDGP:} it was observed that the family of \CY\ manifolds
described by (\ref{eq:Bmod}) admits a symmetry $R$ defined by
$z_0\to\zeta z_0$, $z_i\to z_i$ for $i>0$, and $\psi\to\zeta^{-1}\psi$.
This symmetry establishes an isomorphism between the \CY\ manifolds
at $\psi$ and at $\zeta\psi$, and shows that the correct parameter
for this family is in fact $\psi^{-5}$ rather than $\psi$.
(This parameter can be seen directly by changing coordinates via
$z_0=\psi^{-1}\tilde z$ to give a new defining equation
\begin{equation}
  \psi^{-5}\tilde z^5+z_1^5+z_2^5+z_3^5+z_4^5-5\tilde z z_1z_2z_3z_4=0
		\label{eq:Bmodnew}
\end{equation}
in which the parameter is visibly $\psi^{-5}$.)
When forming a quotient of this family by $\widehat G$ to obtain
the mirror family of $Q$, $R$ remains a symmetry of the quotient---in
fact, $\widehat G$ acts on (\ref{eq:Bmodnew}) equally well as
on (\ref{eq:Bmod}).  However, when forming the quotient by our
group $\widetilde G$ which includes a permutation, $R$ is no
longer a symmetry---it does not normalize the group $\widetilde G$,
nor does it preserve the alternate form (\ref{eq:Bmodnew}) of
the defining equation.

This immediately tells us that the chiral ring of $X$ has ``lost'' some
information concerning the conformal field theory. The chiral ring as
calculated in \cite{CDGP:} is a function of $\psi^5$. However since
$R$ is not a symmetry of $Y$ we expect the points given by
$\psi,\zeta\psi,\zeta^2\psi,\ldots$ to correspond to different
conformal field theories.\footnote{It might be objected that there
could be some other symmetry, not manifest in the present description,
which produces an isomorphism between the theory at $\psi$ and
the theory at $\zeta\psi$.  The calculations in the next section will
demonstrate that this is not the case.}

In order to construct the mirror map between the A- and B-models we
first require a set of ``flat'' coordinates on the B-model moduli
space (see \cite{BCOV:big} for a full discussion of this issue). This
is obtained by considering the variation of Hodge structure on $Y$.
That is, we consider the periods $\varpi_i=\int_{\gamma_i}\Omega$
where $\Omega$ is a (3,0)-form and $\gamma_i$ are elements of
$H_3(Y)$.

Let us use $q$ to denote the image of $e$, the fundamental generator
of $H_2(X)$, under the action of an element of $\Hom(H_2(X),\C^*)$.
The mirror map then relates the A- and B-model moduli spaces by
relating $q$ to $\psi$. The local geometry of the $N$=(2,2) moduli
space tells us that \cite{BCOV:big}
\begin{equation}
  q = \exp\left(\frac{\varpi_1(\psi)}{\varpi_0(\psi)}\right),
\end{equation}
for two suitably chosen periods $\varpi_0$ and $\varpi_1$. To find
exactly which periods to use one must look at global considerations of
$H_3(Y)$ as discussed in \cite{Mor:cp}. Rather than use this method
directly we may use a simple argument as follows. By mapping the
correlation functions of the A-models of $X$ and $Q$ to each other we
obtain a five-fold cover of the moduli space of $Q$ by the moduli
space of $X$. This cover is branched at $q=0$, which represents the
large radius limit.  A path once around this
point in the $X$ moduli space corresponds to changing $B$ by the generator
of $H^2(X,\Z)$.   But since $e_Q$, the generator of $H_2(Q)$, descends
to $5e$ on $X$, a path which winds once around the large radius
limit of the $Q$ moduli space
(changing that space's $B$ by the generator of $H^2(Q,\Z)$)
 will wind 5 times around
$q=0$ in the $X$ moduli space.
Thus, denoting by
$q_Q$ the image of $e_Q$ for the $Q$ moduli space, we obtain
\begin{equation}
  q^5 = q_Q.
\end{equation}
{}From \cite{CDGP:} it then follows that
\begin{equation}
  \eqalign{\varpi_0&=\sum_{N=0}^\infty\frac{(5N)!}{(N!)^5}(5\psi)^{-5N}\cr
  \varpi_1&=-\varpi_0\log(5\psi)+\sum_{N=1}^\infty\frac{(5N)!}{(N!)^5}
  \left[\Psi(5N+1)-\Psi(N+1)\right](5\psi)^{-5N}.\cr}
\end{equation}


\section{Some curve counting} 	\label{s:count}

We may now proceed and count rational curves on $X$. The suitably
normalized three-point function for the A-model is
\begin{equation}
  \langle{\cal O}^3\rangle=25+14375q^5+24384375q^{10}+\ldots,
		\label{eq:rcX}
\end{equation}
It follows \cite{CDGP:,AM:rat} that, for $n_i$ the number of rational
curves of degree $i$ on a generic manifold diffeomorphic to $X$ and
$n_i(Q)$ the same quantity for $Q$, we have
\begin{equation}
  \eqalign{n_i&=0, \qquad\qquad{\rm when}\;5\nmid i,\cr
  n_{5i}&=n_i(Q)/25.\cr}
\end{equation}
This is exactly what we would expect from geometry. The group $G$ is
of order 25 and acts freely on $Q$. Since rational curves do not admit
a freely acting symmetry, $G$ must identify the rational curves on $Q$
in groups of 25 (as observed earlier).
A curve of degree $i$ on $Q$ maps into a curve of
degree $5i$ on $X$ because of the relationship between $H_2$ of the
two spaces.\footnote{The {\em degree}\/ in both cases refers to
the number of intersection points with a generator of $H^2$.}

The form of the expression (\ref{eq:rcX}) shows that the chiral ring of
$X$ does not contain enough information to classify the conformal
field theory. This series may be considered as an instanton expansion.
Since this is a tree-level computation, the instantons are spheres.
Spheres correspond to elements of $\pi_2(X)$ and thus can only
represent homology classes which are a multiple of $5e$. Thus
(\ref{eq:rcX}) is a power series in $q^5$ and cannot fully distinguish
between conformal field theories.

In order to measure any quantity which depends properly on $q$ rather
than $q^5$ we must therefore go beyond tree level. Non trivial
information is obtained beyond genus 0 when the topological A-model is
coupled to gravity \cite{BCOV:ell,BCOV:big}. In this case one may
consider a partition function $F_1$ defined for one-loop world-sheets.
This partition function contains information concerning elliptic
curves on the target space and, with luck, may be used to count them
as follows \cite{BCOV:ell}.
The holomorphic anomaly dictates that $F_1$ is of
the form
\begin{equation}
  F_1 = \log\left[\left(\frac\psi{\varpi_0}\right)^w\!\!
	f(\psi)\,q\frac{d\psi}{dq}\right] + {\rm const},
\end{equation}
where $w=3+h^{1,1}-\chi/12$ (which is
$\ff{14}3$ for $X$), and $f(\psi)$ is an unknown
holomorphic
function of $\psi$. The relationship between $F_1$ and the number,
$n_i$, of rational curves of degree $i$ and elliptic curves, $d_i$, of degree
$i$ on $X$ is given by
\begin{equation}
  F_1 = -\ff1{12}(c_2.e)\log q -\sum_i\left\{2d_i\log\eta(q^i)
   +\ff16n_i\log(1-q^i)\right\}
   + {\rm const},	\label{eq:F1q}
\end{equation}
where
\begin{equation}
  \eta(q) = \prod_{n=1}^\infty(1-q^n),
\end{equation}
and $(c_2.e)$ is obtained by wedging the second Chern class of $X$
with the 2-form dual to $e$ and integrating over $X$.

Knowing $c_2(X)$ and the fact that $F_1$ should be finite for a good
conformal field theory is often sufficient to determine $f(\psi)$.
This was the case for all the examples studied in
\cite{BCOV:ell,CDFKM:I,CFKM:II} but fails for our example. However, we
may find the solution by working a little harder.
The function $f(\psi)$
is generally of the form
\begin{equation}
  f(\psi) = \prod_{s}(\psi_s-\psi)^{a_s},
\end{equation}
where $s$ runs over the points (where $\psi=\psi_s$) in the moduli
space where the conformal field theory is ``bad''. The constants $a_s$
are to be determined. In our example, $Y$ is singular (and thus the
associated conformal field theory is bad) whenever $\psi^5=1$. The
fact that the series part of the the expansion in (\ref{eq:F1q})
has rational coefficients means that $F_1$ must be of the form
\begin{equation}
  f(\psi) = (1-\psi)^{a_0}(1+\psi+\psi^2+\psi^3+\psi^4)^{a_1}.
\end{equation}
The fact that $(c_2.e)=10$ for $X$ the tells us that $a_0+4a_1=-\ff{29}6$.

We can also directly count the number of
degree one elliptic curves on $X$ as follows. In contrast to the usual
cases (see for example \cite{Katz:m}) this may be done very explicitly.
The inverse image of such a curve has total degree $5$ on $Q$.
In principle this might split as $5$ curves of degree 1, permuted by
$G$ (and each invariant under some $\Z_5$ subgroup), but since there
are no elliptic curves of degree 1 on $Q$ this is impossible.
Thus, the inverse image is an irreducible curve of degree 5, preserved
by $G$.  Our task is to count those curves.

Any elliptic curve of degree 5 in $\P^4$ is defined by an ideal generated by
 $5$ quadrics.  When the curve is preserved by $G$, this space of quadrics
must form a projective
representation of $G$, and in fact can be generated by  $5$
quadrics of the form
\begin{equation}
  \eqalign{\alpha z_0^2 + &\beta z_1z_4 + \gamma z_2z_3 \cr
      \alpha z_1^2 + &\beta z_2z_0 + \gamma z_3z_4 \cr
      &\vdots\cr
      \alpha z_4^2 + &\beta z_0z_3 + \gamma z_1z_2\cr}	\label{eq:cvs}
\end{equation}
for some constants $\alpha$, $\beta$, and $\gamma$ (cf.\ \cite{Hulek:}).
Generic values of
those constants lead to 5 quadrics which do not intersect; however,
for any constants satisfying
\begin{equation}
\alpha^2+\beta\gamma=0,
\label{eq:param}
\end{equation}
the intersection is a curve of degree 5.\footnote{This is most easily
seen by intersecting with $z_0=0$ and explicitly solving for the
points of intersection.}  A result from classical projective geometry
(cf.\ \cite{GH:alg}) says that the genus of this curve is at most 1.
On the other hand, if the curve does not pass through the fixed points
of $G$ on $\P^4$, i.e.\ if
\begin{equation}
\alpha\ne0, \text{ and }
\alpha + \zeta\beta + \zeta^{-1}\gamma\ne0 \text{ for all }\zeta^5=1,
\label{eq:condition}
\end{equation}
then $G$ acts without fixed points so the genus cannot be 0 and must
be 1.

When (\ref{eq:param}) is satisfied and $\alpha\ne0$, we can take $\alpha=1$,
$\beta=-1/a$, $\gamma=a$ (as in \cite{Hulek:}).  The corresponding
quadrics (\ref{eq:cvs}) form a Gr\"obner basis for the ideal of
the curve, with leading monomials $z_2z_3$, $z_3z_4$, $z_2^2$, $z_3^2$,
$z_4^2$.  Since this is so, it is easy to find the condition that
the quintic $Q$ defined by (\ref{eq:Fermat}) contain the curve---it is
simply
\begin{equation}
  a^{10}+6a^5-1=0.
\end{equation}
Thus there are precisely 10 such curves.   (Note that all solutions
also satisfy (\ref{eq:condition}).)  These descend to 10 elliptic
curves on $X$, and we conclude that
 $d_1=\ff52(a_1-a_0)=10$. This tells us that $a_0=-\ff{25}6$
and $a_1=-\ff16$.  It also shows, as asserted
earlier, that (\ref{eq:ES3}) does not split.

It is interesting to note that it appears that the exponents $a_s$
might be determined purely by the type of singularities of $Y$ when
$\psi=\psi_s$.
As noted above, $Y$ is singular when $\psi^5=1$. If $\psi\neq1$ then
the form of the singularity is that of a single isolated ``simple''
singularity, i.e., it is locally of the form of the hypersurface
\begin{equation}
  x_1x_2=x_3x_4
\end{equation}
in $\C^4$ near the origin. For $\psi=1$, there are 5 singularities
locally of the form of an orbifold of a simple singularity by $\Z_5$.
In all the cases considered in \cite{BCOV:ell,CDFKM:I,CFKM:II} the
value of $a_s$ for a simple singularity is $-\ff16$ and indeed we have
found the same value in this case.

We now have
\begin{equation}
  F_1 = -\ff56\log q+{\rm const}+20q+50q^2+\ff{500}3q^3+\ldots,
	\label{eq:ecX}
\end{equation}
and we determine
\begin{equation}
  \eqalign{d_1&=10\cr d_2&=10\cr d_3&=70\cr d_4&=280\cr
	d_5&=888\cr &\quad\vdots\cr}	\label{eq:dcount}
\end{equation}
We have extended this calculation through $d_{125}$ and find positive
integers for every degree.


\section{Conclusions} 	\label{s:conc}

We have shown that the space of correlation functions for the A-model
on $X$
does not faithfully represent the moduli space of conformal field
theories on $X$. That is, the chiral ring is not sufficient to
determine the conformal field theory. In this case for generic theories,
there are 5 conformal field theories for each chiral ring. The
exception the this occurs at $\psi=0$ or $\infty$ where the chiral ring
does uniquely determine the theory. This ambiguity is clearly shown in
(\ref{eq:rcX}), valid for small $q$, where the correlation functions are
a function of $q^5$. This occurs because the correlation function is at
tree level and spheres do not span $H_2(X)$. One needs to go to a loop
effect such as (\ref{eq:ecX}) to observe a faithful $q$ dependence. In
our example, $H_2(X)$ is generated by tori and so a one-loop partition
function suffices.

Our results can also be interpreted as providing a
counterexample\footnote{Further analysis of the B-model is likely
needed to make this into a mathematically rigorous
counterexample.}
to the global Torelli problem for \CY\ manifolds (contrary to
assertions in \cite{Tod:WP}).
The Torelli problem asks whether the variation of Hodge structure
determines the complex structure on the manifold.  The variation
of Hodge structure is completely determined by the chiral ring,
and so depends only on $\psi^{-5}$, but we have seen that the conformal
field theory on the manifold (and so presumably the complex structure
itself) actually depends on $\psi$.

If a manifold is simply connected, we have $\pi_2=H_2$ and so $H_2$ is
generated by spheres. This suggests that the failure of the chiral
ring to determine the conformal field theory is caused by non-trivial
$\pi_1$. Indeed it is precisely the (torsion)
group $H^2(\pi_1(X),\C^*)$ which
led to the $\Z_5$ ambiguity in the identification of the conformal
field theory from the chiral ring. One is thus led to hypothesize
that this is precisely the data required to supplement the chiral
ring (cf.~question 6 in \cite{Vafa:vacua}). That is, given a theory
corresponding to target space $X$ with a
mirror $Y$, in the \CY\ phase of both $X$ and $Y$ (i.e., near the large
radius limits) we might have\footnote{Modulo the $\Z_2$ symmetry
coming from complex conjugation.}
\begin{equation}
  \vcenter{\hbox{Moduli Space of}\break\hbox{$N$=(2,2) models}}
	\;\cong\;
  \vcenter{\hbox{Moduli Space}\break\hbox{\hfill of A-ring\hfil}}
  	\;\times\;
  \vcenter{\hbox{Moduli Space}\break\hbox{\hfill of B-ring\hfil}}
  \;\times H^2(\pi_1(X),\C^*)\times H^2(\pi_1(Y),\C^*),
\end{equation}
where by A-ring we mean the chiral ring of the $N$=(2,2) theory as
determined by the A-model and similarly for the B-ring.

One of the motivations for studying this model was to hope to gain
some understanding of the form of the moduli space when there is
torsion in $H_2(X)$ although this did not actually happen for the case
we considered. This might have provided some insight into the
situation when there is ``discrete torsion'' in orbifolds
\cite{Vafa:tor}. Although mirror symmetry has managed to give a
complete picture of the blowing-up of an orbifold without discrete
torsion, at least in the case where points are fixed by abelian groups
(see, for example, \cite{me:orb2}), little is known about this same
process when some discrete torsion is present. Therefore, although
considerable progress has been made in recent years on the form of the
moduli space of $N$=(2,2) theories there is still much to be understood.


\section*{Acknowledgements}

It is a pleasure to thank S. Katz, R. Plesser, C. Vafa, and E. Witten for
useful
conversations.
D.R.M. thanks the School of Natural Sciences, Institute for Advanced
Study for hospitality during the preparation of this paper.
The work of P.S.A. was supported by NSF grant PHYS92-45317,
and the work of D.R.M. was supported by an American Mathematical
Society Centennial Fellowship.


\begin{thebibliography}{10}

\bibitem{CHSW:}
P.~Candelas, G.~Horowitz, A.~Strominger, and E.~Witten,
\newblock {\em Vacuum Configuration for Superstrings},
\newblock Nucl. Phys. {\bf B258} (1985) 46--74.

\bibitem{LVW:}
W.~Lerche, C.~Vafa, and N.~P. Warner,
\newblock {\em Chiral Rings in $N=2$ Superconformal Theories},
\newblock Nucl. Phys. {\bf B324} (1989) 427--474.

\bibitem{W:AB}
E.~Witten,
\newblock {\em Mirror Manifolds and Topological Field Theory},
\newblock in S.-T. Yau, editor, ``Essays on Mirror Manifolds'', International
  Press, 1992.

\bibitem{Vafa:vacua}
C.~Vafa,
\newblock {\em Superstring Vacua},
\newblock in ``1989 Summer School in High Energy Physics and Cosmology,
  Trieste, Italy'', pages 145--177, World Scientific, Singapore, 1989.

\bibitem{N:torus}
K.~S. Narain,
\newblock {\em New Heterotic String Theories in Uncompactified Dimensions $<$
  10},
\newblock Phys. Lett. {\bf 169B} (1986) 41--46.

\bibitem{AM:K3p}
P.~S. Aspinwall and D.~R. Morrison,
\newblock {\em String Theory on K3 Surfaces},
\newblock Duke and IAS 1994 preprint DUK-TH-94-68, IASSNS-HEP-94/23,
  hep-th/9404151,
\newblock to appear in ``Essays on Mirror Manifolds 2''.

\bibitem{DW:gp}
R.~Dijkgraaf and E.~Witten,
\newblock {\em Topological Gauge Theories and Group Cohomology},
\newblock Commun. Math. Phys. {\bf 129} (1990) 393--429.

\bibitem{SW:spin}
N.~Seiberg and E.~Witten,
\newblock {\em Spin Structures in String Theory},
\newblock Nucl. Phys. {\bf B276} (1986) 272--290.

\bibitem{Vafa:tor}
C.~Vafa,
\newblock {\em Modular Invariance and Discrete Torsion on Orbifolds},
\newblock Nucl. Phys. {\bf B273} (1986) 592--606.

\bibitem{W:phase}
E.~Witten,
\newblock {\em Phases of $N=2$ Theories in Two Dimensions},
\newblock Nucl. Phys. {\bf B403} (1993) 159--222.

\bibitem{AGM:II}
P.~S. Aspinwall, B.~R. Greene, and D.~R. Morrison,
\newblock {\em \CY\ Moduli Space, Mirror Manifolds and Spacetime Topology
  Change in String Theory},
\newblock Nucl. Phys. {\bf B416} (1994) 414--480.

\bibitem{BT:}
R.~Bott and L.~W. Tu,
\newblock {\em Differential Forms in Algebraic Topology},
\newblock Springer-Verlag, New York, 1982.

\bibitem{EM:hom}
S.~Eilenberg and S.~MacLane,
\newblock {\em Relations Between Homology and Homotopy Groups of Spaces},
\newblock Ann. Math. {\bf 46} (1945) 480--509.

\bibitem{GP:orb}
B.~R. Greene and M.~R. Plesser,
\newblock {\em Duality in \CY\ Moduli Space},
\newblock Nucl. Phys. {\bf B338} (1990) 15--37.

\bibitem{ALR:coup}
P.~S. Aspinwall, C.~A. L{\"u}tken, and G.~G. Ross,
\newblock {\em Construction and Couplings of Mirror Manifolds},
\newblock Phys. Lett. {\bf 241B} (1990) 373--380.

\bibitem{CDGP:}
P.~Candelas, X.~C. de~la Ossa, P.~S. Green, and L.~Parkes,
\newblock {\em A Pair of Calabi-Yau Manifolds as an Exactly Soluble
  Superconformal Theory},
\newblock Nucl. Phys. {\bf B359} (1991) 21--74.

\bibitem{BCOV:big}
M.~Bershadsky, S.~Cecotti, H.~Ooguri, and C.~Vafa,
\newblock {\em Kodaira-Spencer Theory of Gravity and Exact Results for Quantum
  String Amplitudes},
\newblock Harvard et al 1993 preprint HUTP-93/A025, hep-th/9309140.

\bibitem{Mor:cp}
D.~R. Morrison,
\newblock {\em Compactifications of Moduli Spaces Inspired by Mirror Symmetry},
\newblock in A.~Beauville et~al., editors, ``Journ\'ees de G\'eom\'etrie
  Alg\'ebrique d'Orsay (Juillet 1992)'', volume 218 of Ast\'erisque, pages
  243--271, Soci\'et\'e Math\'ematique de France, 1993.

\bibitem{AM:rat}
P.~S. Aspinwall and D.~R. Morrison,
\newblock {\em Topological Field Theory and Rational Curves},
\newblock Commun. Math. Phys. {\bf 151} (1993) 245--262.

\bibitem{BCOV:ell}
M.~Bershadsky, S.~Cecotti, H.~Ooguri, and C.~Vafa,
\newblock {\em Holomorphic Anomalies in Topological Field Theories},
\newblock Nucl. Phys. {\bf B405} (1993) 279--304.

\bibitem{CDFKM:I}
P.~Candelas, X.~de la Ossa, A.~Font, S.~Katz, and D.~R. Morrison,
\newblock {\em Mirror Symmetry for Two Parameter Models --- I},
\newblock Nucl. Phys. {\bf B416} (1994) 481--562.

\bibitem{CFKM:II}
P.~Candelas, A.~Font, S.~Katz, and D.~R. Morrison,
\newblock {\em Mirror Symmetry for Two Parameter Models --- II},
\newblock Texas et al 1994 preprint UTTG-25-93, hep-th/9403187.

\bibitem{Katz:m}
S.~Katz,
\newblock {\em Rational Curves on Calabi-Yau Threefolds},
\newblock in S.-T. Yau, editor, ``Essays on Mirror Manifolds'', International
  Press, 1992.

\bibitem{Hulek:}
K.~Hulek,
\newblock {\em Projective Geometry of Elliptic Curves}, volume 137 of
  Ast\'erisque,
\newblock Soci\'et\'e Math\'ematique de France, 1986.

\bibitem{GH:alg}
P.~Griffiths and J.~Harris,
\newblock {\em Principles of Algebraic Geometry},
\newblock Wiley-Interscience, 1978.

\bibitem{Tod:WP}
A.~N. Todorov,
\newblock {\em The {W}eil-{P}etersson Geometry of the Moduli Space of
  {$SU(n{\ge}3)$} ({C}alabi-{Y}au) Manifolds, {I}},
\newblock Commun. Math. Phys. {\bf 126} (1989) 325--246.

\bibitem{me:orb2}
P.~S. Aspinwall,
\newblock {\em Resolution of Orbifold Singularities in String Theory},
\newblock IAS 1994 preprint IASSNS-HEP-94/9, hep-th/9403123,
\newblock to appear in ``Essays on Mirror Manifolds 2''.

\end{thebibliography}

\end{document}